\documentclass[aip,apl]{revtex4-1}

\usepackage{graphicx}

\begin{document}

\preprint{}

\title{Charge dynamics of a single donor coupled to a few electrons quantum dot in silicon} 

\author{G. Mazzeo}
\affiliation{Dipartimento di Scienza dei Materiali, Universit\`a degli Studi Milano-Bicocca, Via Cozzi 53, I-20125 Milano, Italy}
\affiliation{Laboratorio MDM, IMM-CNR, Via Olivetti 2, I-20864 Agrate Brianza, Italy }

\author{E. Prati}
\author{M. Belli}
\author{G. Leti}
\author{S. Cocco}
\affiliation{Laboratorio MDM, IMM-CNR, Via Olivetti 2, I-20864 Agrate Brianza, Italy }

\author{M. Fanciulli}
\affiliation{Dipartimento di Scienza dei Materiali, Universit\`a degli Studi Milano-Bicocca, Via Cozzi 53, I-20125 Milano, Italy}
\affiliation{Laboratorio MDM, IMM-CNR, Via Olivetti 2, I-20864 Agrate Brianza, Italy }

\author{F. Guagliardo}
\author{G. Ferrari}
\affiliation{Dipartimento di Ingegneria Elettronica, Politecnico Milano, Milano, Italy}

\date{\today}

\begin{abstract}
We study the charge transfer dynamics between a silicon quantum dot and an individual phosphorous donor using the conduction through the quantum dot as a probe for the donor ionization state. We use a silicon \emph{n}-MOSFET (metal oxide field effect transistor) biased near threshold in the SET regime with two side gates to control both the device conductance and the donor charge. Temperature and magnetic field independent tunneling time is measured. We measure the statistics of the transfer of electrons observed when the ground state $D^0$ of the donor is aligned with the SET states. 


\end{abstract}


\maketitle 

\begin{figure}
\includegraphics{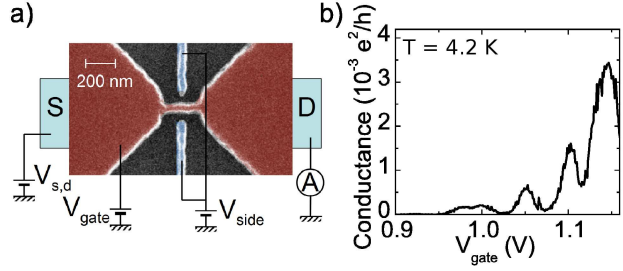}
\caption{\label{fig:device}(color online) a) SEM image of a device nominally identical to the one characterized and schematic representation of the circuit connections. b) Source-drain current vs.\ main gate bias, with the side gates grounded, measured near threshold at 4.2 K.} 
\end{figure}

\begin{figure}
\includegraphics{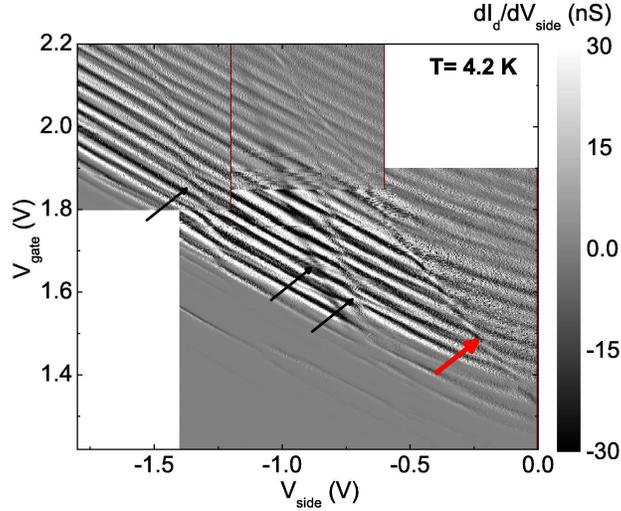}
\caption{\label{fig:IdVgVg2der} (color online) Derivative of the source-drain current vs.\ side gate bias ($dI_d/dV_{side}$) as a function of main gate and side gates polarization at 4.2 K. The arrows indicate the discontinuity lines attributed to donor ionizations. The bold line indicates the region of the donor characterized in this work.} 
\end{figure}

\begin{figure}
\includegraphics{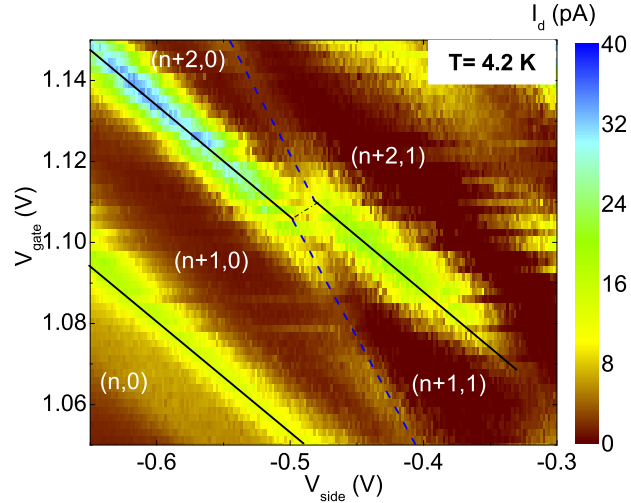}
\caption{\label{fig:IdVgVg2zoom}(color online) Detail of the $I_d$ vs.\ $V_g$ and $V_{side}$ in the quantum dot region near the ionization of one donor acquired at $4.2$ K. The solid and dashed lines represent the charge stability points for the quantum dot and the donor respectively. Current flow is only observed along the quantum dot charge stability line.} 
\end{figure}

\begin{figure}
\includegraphics{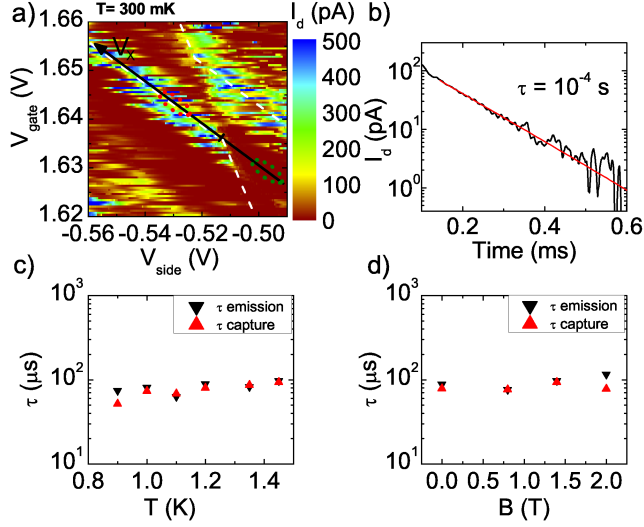}
\caption{\label{fig:300mK}(color online) a) Detail of the scan of the source-drain current as a function of the two gates polarization at 300 mK. The solid arrow represents the line along which the potential is moved for the 1D scans (the arrow indicates the direction of V$_x$ positive. b) Average of the transient quantum dot current acquired after the switching of the bias to allow tunneling of one electron into the donor state. c,d) Dependence of the average tunneling time as a function of the temperature (c) and the magnetic field (d).}
\end{figure}

\begin{figure}
\includegraphics{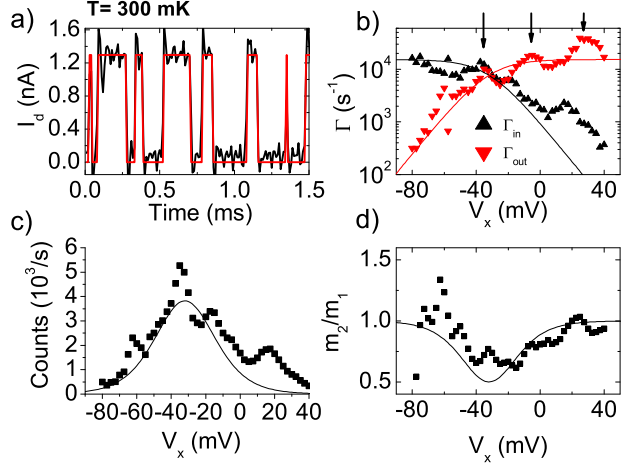}
\caption{\label{fig:rts}(color online) a) Typical RTS trace acquired near the ionization point of the donor and signal reconstructed via a threshold algorithm.  b) Average values for the tunneling rates $\Gamma_{in}$ and $\Gamma_{out}$ as a function of the detuning potential V$_x$ and fitting according to Eq.~\ref{gammainout}. The arrows evidence the fluctuation in the $\Gamma_{out}$ that can be attributed to resonant tunneling to the orbitals in the quantum dot. c,d) Average number of tunneling events per unit time (c) and Fano factor (d) (squares) extracted from the RTS traces. The solid lines represent the theoretical value extracted from the FCS model using the parameters obtained from the fitting of the data in (b).}
\end{figure}

It has been proposed that the manipulation of electron spins bound to single donors represents a solid platform for quantum information processing.\cite{Kane} Building such a system will require a profound understanding of the donor electrons properties, such as spin coherence and decay times, and hyperfine coupling to the nuclear spin environment. While relevant information can be extracted with experimental techniques, such as conventional electrical measurements and electron spin resonance (ESR) spectroscopy, sensitive to the donor in the bulk of the semiconductor, many properties of interest differ when donors are included into nanostructured devices because of the presence of material interfaces,\cite{DonorInNanostructures} confinement in potential wells, and high electric fields.\cite{PStark} The effect of these parameters can only be evidenced in devices where experiments are performed at the single donor level. Recently there has been significant progress in the detection of charge and spin at the single donor level in silicon\cite{MorelloSingleSpinNature} on devices achieving capacitive and tunnel coupling between a few locally implanted donors and a silicon SET working as a charge detector. We have also studied in the past the interaction of single donors in the channel of nanometer scaled silicon MOSFETs (metal oxide field effect transistors) with quantum dot states at low temperature.\cite{PratiAPL} Here, we achieve the transfer of an electron from the quantum dot to a donor in the vicinity of the channel by controlling multiple gates and detect the modification of the charge state by observing the current flowing in the transistor. In this work we improve the capability to detect the charge state of the donor and achieve real time detection of the charge state by employing an \emph{ad-hoc} developed device design. While based on a working principle similar to what proposed in Ref.\ \onlinecite{MorelloPRB}, we focus on the development of a more simple and robust technology employing a single level of metallization and a substrate uniformly doped with phosphorous. The separate control of multiple gates allow biasing the device in a region where the ionization of only one donor is energetically allowed. From time averaged traces of the capture and emission process we measure the average tunneling time that is observed to be independent on both temperature and magnetic field. From real time traces of the charge transfer we measure the average tunneling rates and the process statistics in the framework of the full counting statistics (FCS) following the model introduced in Ref.\ \onlinecite{levitovFCS} and developed in Ref.\ \onlinecite{BagretsFCS}. The tunneling rates to and from the quantum dot reflect the internal energy level spectrum of the quantum dot and its occupation. As predicted, the capture process is measured to be sub-poissonian as required by the Coulomb blockade regime that allows the capture of only one electron in the donor.
\newline
The experiments were performed on a silicon \emph{n}-MOSFET with an engineered structure of gates. The device was realized with a standard process however, in contrast to what used for the realization of silicon quantum dots in other experiments,\cite{MorelloSingleSpinNature,SandiaMOSChargeSensing,JiangMOST1} the functionalities were obtained with only one level of metalization. A scanning electron microscope (SEM) image of a device nominally equal to the one here reported is shown in Figure \ref{fig:device}(a). 
The main gate is shaped to form the $W = 40$ nm and $L = 300$ nm channel of the transistor when a sufficiently positive gate voltage $V_g$ is applied. Near the thin channel two side gates are realized at $60$ nm from the main gate. The two side gates are biased at the same potential $V_{side}$ during the experiments. The gate oxide thickness, $35$ nm, is chosen to guarantee that both main and side gates achieve a good control on the charge in the channel. This thin channel transistor is used as a highly sensitive electrometer to detect the charge state of nearby donors. The phosphorus donors interacting with the channel of the transistor are provided as a uniform doping distribution in the substrate of $1 \div 5 \times 10^{15}$  cm$^{-3}$ and a total of $1 \div 10$ donors are expected to be near the active region of the device. At the low operating temperature of the device all donors in the substrate are in the neutral charge state (D$^{0}$) and do not contribute to the source/drain conduction. Preliminary characterization was performed in liquid helium at $4.2$ K while time transient experiments were performed in a $^3$He refrigerator at $300$ mK. The sample was connected to a cryogenic amplifier thermally anchored to the $1$~K-pot of the cryostat. The amplifier is a custom circuit realized in standard $3.3$ V complementary metal oxide semiconductor (CMOS) technology. With this configuration the SET current could be measured with a $34$ kHz bandwidth and $2$ pA noise at full bandwidth.
\newline
By exploiting the disorder at the interface, causing the formation of isolated potential minima, when biased near threshold the transistor behaves as a single electron quantum dot as evidenced by the drain current ($I_d$) vs.\ the main gate voltage ($V_g$) trace at $4.2$ K reported in Figure \ref{fig:device}(b). When the electron density in the channel is increased the potential fluctuations are screened and the ohmic behavior is recovered. 
Measuring the $I_d$ vs.\ $V_g$ and $V_{side}$ (side gate bias) and plotting the derivative of the current to the side gate bias, as reported in Figure \ref{fig:IdVgVg2der}, 
it is shown how both gates control the transistor current. Plotting the derivative of the current evidences the Coulomb blockade peaks that show a fairly regular pattern characterized by parallel conductance lines.
Along several lines in the $d(I_d)/d(V_{side})$ vs.\ $(V_{g},V_{side})$ diagram of Figure \ref{fig:IdVgVg2der}, evidenced by the arrows, it is possible to observe discontinuities in the otherwise regular pattern of the Coulomb blockade peaks. These discontinuities are associated with a sudden modification of the potential of the quantum dot caused by the ionization/charging of a defect state in the vicinity of the channel, as also reported in Ref.\ \onlinecite{MorelloSingleSpinNature} and Ref.\ \onlinecite{SimmonsPChargeSensing}. The lines evidenced in Figure \ref{fig:IdVgVg2der} are characterized by different slopes in the $V_{g},V_{side}$ plane and are thus associated to distinct defects with different coupling to the two gates. After each cool down cycle we observed a shift of the Coulomb blockade features in the transport through the quantum dot that are attributed to defects at the $Si/SiO_2$ interface; however the slope and relative position of the features attributed to the donors ionization were reproducible. This allowed an unambiguous identification of the donors and, after each cool down cycle, is was possible to tune the device to characterize always the same donor. This required an adjustment of the bias point. The bold arrow in Figure \ref{fig:IdVgVg2der} evidences the feature associated to the donor that has been characterized in detail.
The features attributed to the charging of a single donor could not be observed on similar devices realized on \emph{p}-type silicon substrates. Both types of devices are based on a \emph{n}-FET and are operated in the same bias range; consequently in both cases the channel interacts with defects with energy states close to the conduction band minimum and would show similar features if those were consequence of the trapping at defect states.
\newline
When the device is polarized in the quantum dot regime the change of the ionization state of a single donor can cause the complete blockade of the source/drain current. This situation is reported in the detail of the 2D gate scan acquired at $4.2$ K in Figure \ref{fig:IdVgVg2zoom}
where the stability lines for the quantum dot and the donor are evidenced. The capacitive coupling between the donor and the quantum dot causes the shift of the quantum dot conduction line bringing the quantum dot out of the resonance line.
\newline
The lever arm factors to the quantum dot, $\alpha_{QD,gate(side)}=d\mu_{QD}/dV_{g(side)}$, have been extracted near the bias point of Figure \ref{fig:300mK}(a) at the temperature of $300$ mK from transport experiments (not shown here), while the arm factors to the donor, $\alpha_{donor,gate(side)}=d\mu_{donor}/dV_{g(side)}$, were extracted from the thermal broadening of the ionization transition, assuming that the quantum dot has a quasi continuous distribution of states with an occupation given from the Fermi distribution, as described in Ref.\ \onlinecite{MorelloSingleSpinNature}. The extracted values are reported in Table~\ref{tab:armfactors} and are compatible with the interpretation of the presence of a donor that is spatially separated from the quantum dot and more closely coupled to the side gates than to the main gate. 
\begin{table} 
\caption{\label{tab:armfactors} Lever arm factors to the quantum dot and the donor from the main and side gates (in eV/V).}
\begin{ruledtabular}
\begin{tabular}{l c c }
\hline  & Main Gate & Side Gate \\ 
\hline Quantum Dot & 0.16 & 0.075 \\ 
\hline Donor & 0.028 & 0.059 \\ 
\hline 
\end{tabular}
\end{ruledtabular}
\end{table}
In all the following experiments, we move the potential along one line on the $V_{g},V_{side}$ diagram, specified by the arrow in Figure \ref{fig:300mK}(a),
and parallel to one of the Coulomb blockade peaks. The origin of the new axis is set so that conduction through the quantum dot is only allowed when the donor is ionized: in this condition the average current through the quantum dot is proportional to the average occupancy of the donor. The new V$_x$ bias is proportional to the energy detuning, $\Delta E$, between the quantum dot and the donor, while moving along this axis keeps the quantum dot chemical potential fixed. The lever arm factor to the donor is  $\alpha_{QD,x} = \Delta E / V_x = (0.009 \pm 0.001)$ eV/V. 
\newline 
The average tunneling times to and from the donor at large bias detuning ($|\Delta E| \gg k_BT$) are measured by switching the bias to allow the capture and release of one electron and measuring the current through the quantum dot. Averaging several acquisitions the exponential decay of the current, and thus of the donor occupancy, is reconstructed, as reported in Figure~\ref{fig:300mK}(b). The decay time extracted from the single exponential fit of the current is reported in Figure~\ref{fig:300mK}(c) and Figure \ref{fig:300mK}(d), and it results to be largely independent on both temperature and magnetic field, proving that transfer of the electron is through elastic tunneling between states with the same spin projection along the magnetic field direction.
\newline 
When the detuning energy $\Delta E$ is comparable with $k_BT$ electrons are continuously exchanged between the quantum dot and the donor. The quantum dot current shows random telegraph signal (RTS) \cite{PratiRTS} reflecting the instantaneous occupation of the donor.. Typical traces and the reconstructed signal are reported in Figure~\ref{fig:rts}(a).
The average time to enter(exit) the donor is related to the tunneling time and the density of available occupied(empty) states for tunneling from(in) the quantum dot. Assuming that the quantum dot presents a quasi continuous density of states occupied according to a Fermi distribution, the tunneling times can be written as in Ref. \onlinecite{QD_SuperPoissonianNoise}:
\begin{equation}\label{gammainout}
\Gamma_{out/in}=\Gamma_0\frac{1}{1+exp(\pm\Delta E / k_BT)}.
\end{equation}
By sweeping the V$_x$ control voltage, we vary $\Delta E$ and the relative occupancies of the empty (high current) and occupied (current blocked) states. The experimental tunneling rates, together with a fit according to Eq.~\ref{gammainout} are reported in Figure~\ref{fig:rts}(b). The value for $\Gamma_0$ corresponds to the tunneling rate that has been measured with the time averaged technique. $\Gamma_0=1.5 \times 10^{4}$ s$^{-1}$. The fit of the data in Figure~\ref{fig:rts}(b) reveals a temperature of the electron system of $T=1.2$ K, higher than the refrigerator temperature of 300 mK. Because of the weak electron-phonon coupling the electron temperature is generally different from the nominal temperature of the $^3$He bath and of the crystal: this has been observed both in our system on different devices\cite{Prati_RTS_Temp,Prati_PRB_triplet_RTS} as well as in other systems.\cite{MorelloSingleSpinNature} The oscillation of the value of the tunnel rates, evidenced by the arrows in Figure~\ref{fig:rts}(b), can be attributed to the discrete spectrum of energy levels in the quantum dot: from the period of the oscillation and the previously determined lever arm factor the energy separation can be roughly estimated to be $300$ $\mu$eV. In a harmonic potential well $\Delta E=\hbar^2m_t^{*-1}l^{-2}$ where $m_t^*$ is the effective transverse electron mass and $l$ is the wavefunction extension. From this equation we obtain an effective dot size of $36$ nm. Because of the confining potential of the dot is induced by charge defect at the $Si/SiO_2$ interface this number should be compared with the expected defect density. A 36 nm linear distance between defects would correspond to an areal density of $8 \times 10^{10}$ cm$^{-2}$ which is compatible with the technology used.
\newline
From the real time traces we also extract the first and second order statistics of the tunneling process. The analysis, in the framework of the full counting statistics, is performed by measuring the probability function of finding \emph{n} transitions in a given time $t_0$ and calculating the first and second order moments as in Ref.\ \onlinecite{QD_SuperPoissonianNoise}: 
\begin{eqnarray}\label{moments}
&&m_1 = <n>_t\\
&&m_2 = <(n-<n>)^2>_t
\end{eqnarray}
where $<$...$>_t$ represents the average over time. The first moment $m_1$ of the probability distribution corresponds to the average number of transitions per unit time, while the second moment $m_2$ is the variance of the probability distribution. In a perfectly Poissonian process, with no correlation between the occurrence of two subsequent events, the Fano factor $F=m_2/m_1$, is always unitary. This value can be modified by the occurrence of some degree of correlation. One notable case is that of electrons tunneling in and out of quantum dots in which, due to the Coulomb blockade, only one electron at a time can be confined.\cite{BagretsFCS} In such systems the Fano factor depends on $\Delta E$ as:\cite{JiangMazzeoRTS}
\begin{equation}\label{Fano}
\frac{m_2}{m_1}=\left(1+\beta^2\right)/2
\end{equation}
where $\beta=2f\left(\Delta E/kT\right)-1$ and $f\left(x\right)$ is the Fermi-Dirac distribution. This value reaches its minimum at $\Delta E=0$, where $\tau_e=\tau_c$ and $F=m_2/m_1=0.5$. In Figure \ref{fig:rts}c and \ref{fig:rts}d we report the values for $m_1$ and the Fano factor extracted from the reconstructed RTS traces together with the values calculated from Eq. \ref{Fano} with the same parameters extracted from the fit of the tunneling rates.
\newline
Concluding, we have analyzed the dynamic charge transfer between a quantum dot and a single donor atom in a silicon device. The real time occupancy of the donor was extracted from the current flowing through the quantum dot. The device, consisting in a silicon \emph{n}-MOSFET, allowed the real time measurement of the charge state of neighboring donors. We studied the interaction with one of the donors with tunneling time in the order of $1 \times 10^{-4}$ s. The tunneling rate is found to be independent on the temperature, indicating elastic tunneling processes. RTS traces were measured near the energy equilibrium point, where electrons are continuously exchanged between the quantum dot and the donor. The tunneling rates to and from the quantum dot reflect the thermally broadened occupation of the energy levels in the quantum dot. As predicted the capture process is measured to be sub-poissonian as required by the Coulomb blockade regime that allows the capture of only one electron in the donor.

\begin{acknowledgments}
This work was partly supported by the CARIPLO Foundation national project ELIOS. 
\end{acknowledgments}


\begin{thebibliography}{10}%
\makeatletter
\providecommand \@ifxundefined [1]{%
 \ifx #1\undefined \expandafter \@firstoftwo
 \else \expandafter \@secondoftwo
\fi
}%
\providecommand \@ifnum [1]{%
 \ifnum #1\expandafter \@firstoftwo
 \else \expandafter \@secondoftwo
\fi
}%
\providecommand \enquote [1]{``#1''}%
\providecommand \bibnamefont  [1]{#1}%
\providecommand \bibfnamefont [1]{#1}%
\providecommand \citenamefont [1]{#1}%
\providecommand\href[0]{\@sanitize\@href}%
\providecommand\@href[1]{\endgroup\@@startlink{#1}\endgroup\@@href}%
\providecommand\@@href[1]{#1\@@endlink}%
\providecommand \@sanitize [0]{\begingroup\catcode`\&12\catcode`\#12\relax}%
\@ifxundefined \pdfoutput {\@firstoftwo}{%
 \@ifnum{\z@=\pdfoutput}{\@firstoftwo}{\@secondoftwo}%
}{%
 \providecommand\@@startlink[1]{\leavevmode}%
 \providecommand\@@endlink[0]{}%
}{%
 \providecommand\@@startlink[1]{%
  \leavevmode
  \pdfstartlink
   attr{/Border[0 0 1 ]/H/I/C[0 1 1]}%
   user{/Subtype/Link/A<</Type/Action/S/URI/URI(#1)>>}%
  \relax
 }%
 \providecommand\@@endlink[0]{\pdfendlink}%
}%
\providecommand \url  [0]{\begingroup\@sanitize \@url }%
\providecommand \@url [1]{\endgroup\@href {#1}{\urlprefix}}%
\providecommand \urlprefix [0]{URL }%
\providecommand \Eprint[0]{\href }%
\@ifxundefined \urlstyle {%
  \providecommand \doi [1]{doi:\discretionary{}{}{}#1}%
}{%
  \providecommand \doi [0]{doi:\discretionary{}{}{}\begingroup
  \urlstyle{rm}\Url }%
}%
\providecommand \doibase [0]{http://dx.doi.org/}%
\providecommand \Doi[1]{\href{\doibase#1}}%
\providecommand \selectlanguage [0]{\@gobble}%
\providecommand \bibinfo [0]{\@secondoftwo}%
\providecommand \bibfield [0]{\@secondoftwo}%
\providecommand \translation [1]{[#1]}%
\providecommand \BibitemOpen[0]{}%
\providecommand \bibitemStop [0]{}%
\providecommand \bibitemNoStop [0]{.\EOS\space}%
\providecommand \EOS [0]{\spacefactor3000\relax}%
\providecommand \BibitemShut [1]{\csname bibitem#1\endcsname}%
\bibitem{Kane}%
  \BibitemOpen
  \bibfield{author}{%
  \bibinfo {author} {\bibfnamefont{B.~E.}\ \bibnamefont{Kane}},\ }%
  \bibfield{journal}{%
  \bibinfo {journal} {Nature}\ }%
  \textbf{\bibinfo {volume} {393}},\ \bibinfo {pages} {133} (\bibinfo {month}
  {May}\ \bibinfo {year} {1998})\BibitemShut{NoStop}%
\bibitem{DonorInNanostructures}%
  \BibitemOpen
  \bibfield{author}{%
  \bibinfo {author} {\bibfnamefont{M.}~\bibnamefont{Diarra}}, \bibinfo {author}
  {\bibfnamefont{Y.-M.}\ \bibnamefont{Niquet}}, \bibinfo {author}
  {\bibfnamefont{C.}~\bibnamefont{Delerue}},\ and\ \bibinfo {author}
  {\bibfnamefont{G.}~\bibnamefont{Allan}},\ }%
  \bibfield{journal}{%
  \bibinfo {journal} {Phys. Rev. B}\ }%
  \textbf{\bibinfo {volume} {75}},\ \bibinfo {pages} {045301} (\bibinfo {month}
  {Jan}\ \bibinfo {year} {2007})\BibitemShut{NoStop}%
\bibitem{PStark}%
  \BibitemOpen
  \bibfield{author}{%
  \bibinfo {author} {\bibfnamefont{M.}~\bibnamefont{Friesen}},\ }%
  \bibfield{journal}{%
  \bibinfo {journal} {Phys. Rev. Lett.}\ }%
  \textbf{\bibinfo {volume} {94}},\ \bibinfo {pages} {186403} (\bibinfo {month}
  {May}\ \bibinfo {year} {2005})\BibitemShut{NoStop}%
\bibitem{MorelloSingleSpinNature}%
  \BibitemOpen
  \bibfield{author}{%
  \bibinfo {author} {\bibfnamefont{A.}~\bibnamefont{Morello}}, \bibinfo
  {author} {\bibfnamefont{J.~J.}\ \bibnamefont{Pla}}, \bibinfo {author}
  {\bibfnamefont{F.~A.}\ \bibnamefont{Zwanenburg}}, \bibinfo {author}
  {\bibfnamefont{K.~W.}\ \bibnamefont{Chan}}, \bibinfo {author}
  {\bibfnamefont{K.~Y.}\ \bibnamefont{Tan}}, \bibinfo {author}
  {\bibfnamefont{H.}~\bibnamefont{Huebl}}, \bibinfo {author}
  {\bibfnamefont{M.}~\bibnamefont{Mottonen}}, \bibinfo {author}
  {\bibfnamefont{C.~D.}\ \bibnamefont{Nugroho}}, \bibinfo {author}
  {\bibfnamefont{C.}~\bibnamefont{Yang}}, \bibinfo {author}
  {\bibfnamefont{J.~A.}\ \bibnamefont{van Donkelaar}}, \bibinfo {author}
  {\bibfnamefont{A.~D.~C.}\ \bibnamefont{Alves}}, \bibinfo {author}
  {\bibfnamefont{D.~N.}\ \bibnamefont{Jamieson}}, \bibinfo {author}
  {\bibfnamefont{C.~C.}\ \bibnamefont{Escott}}, \bibinfo {author}
  {\bibfnamefont{L.~C.~L.}\ \bibnamefont{Hollenberg}}, \bibinfo {author}
  {\bibfnamefont{R.~G.}\ \bibnamefont{Clark}},\ and\ \bibinfo {author}
  {\bibfnamefont{A.~S.}\ \bibnamefont{Dzurak}},\ }%
  \bibfield{journal}{%
  \bibinfo {journal} {Nature}\ }%
  \textbf{\bibinfo {volume} {467}},\ \bibinfo {pages} {687} (\bibinfo {year}
  {2010})\BibitemShut{NoStop}%
\bibitem{PratiAPL}%
  \BibitemOpen
  \bibfield{author}{%
  \bibinfo {author} {\bibfnamefont{E.}~\bibnamefont{Prati}}, \bibinfo {author}
  {\bibfnamefont{M.}~\bibnamefont{Belli}}, \bibinfo {author}
  {\bibfnamefont{S.}~\bibnamefont{Cocco}}, \bibinfo {author}
  {\bibfnamefont{G.}~\bibnamefont{Petretto}},\ and\ \bibinfo {author}
  {\bibfnamefont{M.}~\bibnamefont{Fanciulli}},\ }%
  \bibfield{journal}{%
  \bibinfo {journal} {Applied Physics Letters}\ }%
  \textbf{\bibinfo {volume} {98}},\ \bibinfo {pages} {053109 } (\bibinfo
  {month} {jan}\ \bibinfo {year} {2011})\BibitemShut{NoStop}%
\bibitem{MorelloPRB}%
  \BibitemOpen
  \bibfield{author}{%
  \bibinfo {author} {\bibfnamefont{A.}~\bibnamefont{Morello}}, \bibinfo
  {author} {\bibfnamefont{C.~C.}\ \bibnamefont{Escott}}, \bibinfo {author}
  {\bibfnamefont{H.}~\bibnamefont{Huebl}}, \bibinfo {author}
  {\bibfnamefont{L.~H.}\ \bibnamefont{Willems~van Beveren}}, \bibinfo {author}
  {\bibfnamefont{L.~C.~L.}\ \bibnamefont{Hollenberg}}, \bibinfo {author}
  {\bibfnamefont{D.~N.}\ \bibnamefont{Jamieson}}, \bibinfo {author}
  {\bibfnamefont{A.~S.}\ \bibnamefont{Dzurak}},\ and\ \bibinfo {author}
  {\bibfnamefont{R.~G.}\ \bibnamefont{Clark}},\ }%
  \bibfield{journal}{%
  \bibinfo {journal} {Phys. Rev. B}\ }%
  \textbf{\bibinfo {volume} {80}},\ \bibinfo {pages} {081307} (\bibinfo {month}
  {Aug}\ \bibinfo {year} {2009})\BibitemShut{NoStop}%
\bibitem{levitovFCS}%
  \BibitemOpen
  \bibfield{author}{%
  \bibinfo {author} {\bibfnamefont{L.~S.}\ \bibnamefont{Levitov}}, \bibinfo
  {author} {\bibfnamefont{H.}~\bibnamefont{Lee}},\ and\ \bibinfo {author}
  {\bibfnamefont{G.~B.}\ \bibnamefont{Lesovik}},\ }%
  \bibfield{journal}{%
  \bibinfo {journal} {Journal of Mathematical Physics}\ }%
  \textbf{\bibinfo {volume} {37}},\ \bibinfo {pages} {4845} (\bibinfo {year}
  {1996})\BibitemShut{NoStop}%
\bibitem{BagretsFCS}%
  \BibitemOpen
  \bibfield{author}{%
  \bibinfo {author} {\bibfnamefont{D.~A.}\ \bibnamefont{Bagrets}}\ and\
  \bibinfo {author} {\bibfnamefont{Y.~V.}\ \bibnamefont{Nazarov}},\ }%
  \bibfield{journal}{%
  \bibinfo {journal} {Phys. Rev. B}\ }%
  \textbf{\bibinfo {volume} {67}},\ \bibinfo {pages} {085316} (\bibinfo {month}
  {Feb}\ \bibinfo {year} {2003})\BibitemShut{NoStop}%
\bibitem{SandiaMOSChargeSensing}%
  \BibitemOpen
  \bibfield{author}{%
  \bibinfo {author} {\bibfnamefont{E.~P.}\ \bibnamefont{Nordberg}}, \bibinfo
  {author} {\bibfnamefont{H.~L.}\ \bibnamefont{Stalford}}, \bibinfo {author}
  {\bibfnamefont{R.}~\bibnamefont{Young}}, \bibinfo {author}
  {\bibfnamefont{G.~A.~T.}\ \bibnamefont{Eyck}}, \bibinfo {author}
  {\bibfnamefont{K.}~\bibnamefont{Eng}}, \bibinfo {author}
  {\bibfnamefont{L.~A.}\ \bibnamefont{Tracy}}, \bibinfo {author}
  {\bibfnamefont{K.~D.}\ \bibnamefont{Childs}}, \bibinfo {author}
  {\bibfnamefont{J.~R.}\ \bibnamefont{Wendt}}, \bibinfo {author}
  {\bibfnamefont{R.~K.}\ \bibnamefont{Grubbs}}, \bibinfo {author}
  {\bibfnamefont{J.}~\bibnamefont{Stevens}}, \bibinfo {author}
  {\bibfnamefont{M.~P.}\ \bibnamefont{Lilly}}, \bibinfo {author}
  {\bibfnamefont{M.~A.}\ \bibnamefont{Eriksson}},\ and\ \bibinfo {author}
  {\bibfnamefont{M.~S.}\ \bibnamefont{Carroll}},\ }%
  \bibfield{journal}{%
  \bibinfo {journal} {Applied Physics Letters}\ }%
  \textbf{\bibinfo {volume} {95}},\ \bibinfo {pages} {202102} (\bibinfo {year}
  {2009})\BibitemShut{NoStop}%
\bibitem{JiangMOST1}%
  \BibitemOpen
  \bibfield{author}{%
  \bibinfo {author} {\bibfnamefont{M.}~\bibnamefont{Xiao}}, \bibinfo {author}
  {\bibfnamefont{M.~G.}\ \bibnamefont{House}},\ and\ \bibinfo {author}
  {\bibfnamefont{H.~W.}\ \bibnamefont{Jiang}},\ }%
  \bibfield{journal}{%
  \bibinfo {journal} {Phys. Rev. Lett.}\ }%
  \textbf{\bibinfo {volume} {104}},\ \bibinfo {pages} {096801} (\bibinfo
  {month} {Mar}\ \bibinfo {year} {2010})\BibitemShut{NoStop}%
\bibitem{SimmonsPChargeSensing}%
  \BibitemOpen
  \bibfield{author}{%
  \bibinfo {author} {\bibfnamefont{S.}~\bibnamefont{Mahapatra}}, \bibinfo
  {author} {\bibfnamefont{H.}~\bibnamefont{B\"uch}},\ and\ \bibinfo {author}
  {\bibfnamefont{M.~Y.}\ \bibnamefont{Simmons}},\ }%
  \bibfield{journal}{%
  \bibinfo {journal} {Nano Letters}\ }%
  \textbf{\bibinfo {volume} {11}},\ \bibinfo {pages} {4376} (\bibinfo {year}
  {2011})\BibitemShut{NoStop}%
\bibitem{PratiRTS}%
  \BibitemOpen
  \bibfield{author}{%
  \bibinfo {author} {\bibfnamefont{E.}~\bibnamefont{Prati}}, \bibinfo {author}
  {\bibfnamefont{M.}~\bibnamefont{Fanciulli}}, \bibinfo {author}
  {\bibfnamefont{G.}~\bibnamefont{Ferrari}},\ and\ \bibinfo {author}
  {\bibfnamefont{M.}~\bibnamefont{Sampietro}},\ }%
  \bibfield{journal}{%
  \Doi{10.1063/1.2939272}{\bibinfo {journal} {Journal of Applied Physics}}\ }%
  \textbf{\bibinfo {volume} {103}},\ \bibinfo {eid} {123707} (\bibinfo {year}
  {2008}),\
  \url{http://link.aip.org/link/?JAP/103/123707/1}\BibitemShut{NoStop}%
\bibitem{QD_SuperPoissonianNoise}%
  \BibitemOpen
  \bibfield{author}{%
  \bibinfo {author} {\bibfnamefont{S.}~\bibnamefont{Gustavsson}}, \bibinfo
  {author} {\bibfnamefont{R.}~\bibnamefont{Leturcq}}, \bibinfo {author}
  {\bibfnamefont{B.}~\bibnamefont{Simovi\ifmmode~\check{c}\else \v{c}\fi{}}},
  \bibinfo {author} {\bibfnamefont{R.}~\bibnamefont{Schleser}}, \bibinfo
  {author} {\bibfnamefont{P.}~\bibnamefont{Studerus}}, \bibinfo {author}
  {\bibfnamefont{T.}~\bibnamefont{Ihn}}, \bibinfo {author}
  {\bibfnamefont{K.}~\bibnamefont{Ensslin}}, \bibinfo {author}
  {\bibfnamefont{D.~C.}\ \bibnamefont{Driscoll}},\ and\ \bibinfo {author}
  {\bibfnamefont{A.~C.}\ \bibnamefont{Gossard}},\ }%
  \bibfield{journal}{%
  \bibinfo {journal} {Phys. Rev. B}\ }%
  \textbf{\bibinfo {volume} {74}},\ \bibinfo {pages} {195305} (\bibinfo {month}
  {Nov}\ \bibinfo {year} {2006})\BibitemShut{NoStop}%
\bibitem{Prati_RTS_Temp}%
  \BibitemOpen
  \bibfield{author}{%
  \bibinfo {author} {\bibfnamefont{E.}~\bibnamefont{Prati}}, \bibinfo {author}
  {\bibfnamefont{M.}~\bibnamefont{Belli}}, \bibinfo {author}
  {\bibfnamefont{M.}~\bibnamefont{Fanciulli}},\ and\ \bibinfo {author}
  {\bibfnamefont{G.}~\bibnamefont{Ferrari}},\ }%
  \bibfield{journal}{%
  \bibinfo {journal} {Applied Physics Letters}\ }%
  \textbf{\bibinfo {volume} {96}},\ \bibinfo {pages} {113109} (\bibinfo {year}
  {2010})\BibitemShut{NoStop}%
\bibitem{Prati_PRB_triplet_RTS}%
  \BibitemOpen
  \bibfield{author}{%
  \bibinfo {author} {\bibfnamefont{E.}~\bibnamefont{Prati}}, \bibinfo {author}
  {\bibfnamefont{M.}~\bibnamefont{Fanciulli}}, \bibinfo {author}
  {\bibfnamefont{G.}~\bibnamefont{Ferrari}},\ and\ \bibinfo {author}
  {\bibfnamefont{M.}~\bibnamefont{Sampietro}},\ }%
  \bibfield{journal}{%
  \bibinfo {journal} {Phys. Rev. B}\ }%
  \textbf{\bibinfo {volume} {74}},\ \bibinfo {pages} {033309} (\bibinfo {month}
  {Jul}\ \bibinfo {year} {2006})\BibitemShut{NoStop}%
\bibitem{JiangMazzeoRTS}%
  \BibitemOpen
  \bibfield{author}{%
  \bibinfo {author} {\bibfnamefont{X.~C.}\ \bibnamefont{Zhang}}, \bibinfo
  {author} {\bibfnamefont{G.}~\bibnamefont{Mazzeo}}, \bibinfo {author}
  {\bibfnamefont{A.}~\bibnamefont{Brataas}}, \bibinfo {author}
  {\bibfnamefont{M.}~\bibnamefont{Xiao}}, \bibinfo {author}
  {\bibfnamefont{E.}~\bibnamefont{Yablonovitch}},\ and\ \bibinfo {author}
  {\bibfnamefont{H.~W.}\ \bibnamefont{Jiang}},\ }%
  \bibfield{journal}{%
  \bibinfo {journal} {Phys. Rev. B}\ }%
  \textbf{\bibinfo {volume} {80}},\ \bibinfo {pages} {035321} (\bibinfo {month}
  {Jul}\ \bibinfo {year} {2009})\BibitemShut{NoStop}%
\end{thebibliography}

\providecommand{\noopsort}[1]{}\providecommand{\singleletter}[1]{#1}%

\pagebreak[4]

\end{document}